# Effect of Indium doping on the superconductivity of layered oxychalcogenide $La_2O_2Bi_3Ag_{1-x}In_xS_6$

Rajveer Jha[1]*, Yosuke Goto[1], Tatsuma D Matsuda[1], Yuji Aoki[1] and Yoshikazu Mizuguchi[1]

[1]Department of Physics, Tokyo Metropolitan University, 1-1 Minami-Osawa, Hachioji, Tokyo 192-0397, Japan

E-mail for corresponding author: *rajveerjha@gmail.com

**Abstract.** We report on the substitution effect of Indium (In) at the Ag site of layered oxychalcogenide $La_2O_2Bi_3Ag_{1-x}In_xS_6$. The $T_c$ decreases with increasing In concentration. A hump in the normal state resistivity at an anomaly temperature ($T^*$) near 180 K was observed for all the samples. The anomaly in the resistivity at $T^*$ is indicating the possible occurrence of a charge-density-wave (CDW) transition. The $T^*$ does not markedly change by In doping. The $x$ dependence of Seebeck coefficient suggests that carrier concentration does not change by In doping. The EDX analysis indicates small amount of Bi deficiency, which suggests that the Bi site is slightly substituted by In. The CDW transition is robust against the In substitution at Ag site, while $T_c$ is decreasing due to the Bi site substitution by In. On the basis of those analyses, we propose that the suppression of superconductivity in the In-doped $La_2O_2Bi_3Ag_{1-x}In_xS_6$ system is caused by negative in-plane chemical pressure effect and partial substitution of In for the in-plane Bi site.

## 1. Introduction
The discovery of superconductivity in layered Bismuth disulphide ($BiS_2$–based) compounds $Bi_4O_4S_3$ with a transition temperature ($T_c$) of 4.5 K [1], $REO_{1-x}F_xBiS_2$ (RE = La, Ce, Nd, Yb, Pr) and $Sr_{1-x}La_xFBiS_2$ with $T_c$ of 2-5 K [1–11] have created tremendous interests of condensed matter physics scientific community [1,2]. The highest $T_c$ up to 11 K was observed in a high-pressure phase of $LaO_{1-x}F_xBiS_2$ [12-16]. Recently, we reported superconductivity in the oxychalcogenide $La_2O_2Bi_3AgS_6$ with $T_c$ of 0.5 K [17]. Furthermore, we reported that $T_c$ can be increased up to 2.4 K by Tin (Sn) substitution at the silver (Ag) site of $La_2O_2Bi_3AgS_6$ [18]. Basically, $La_2O_2Bi_3AgS_6$ has a crystal structure similar to that of $LaOPbBiS_3$, which is composed of $La_2O_2$ blocking layer and $Bi_2Pb_2S_6$ (or $Bi_3AgS_6$ for $La_2O_2Bi_3AgS_6$) conducting layer [19-21]. In those oxychalcogenide family ($La_2O_2M_4S_6$: $M$ = metal), the NaCl-type layer ($M_2S_2$ layer) is inserted into the van der Waals gap of two $BiS_2$ layers [19]. The other factor essential for the superconductivity in the $BiS_2$ family is the chemical pressure effect, which suppresses in-plan disorder [22-23]. Therefore, we aimed to study the further substitution effects for the $La_2O_2Bi_3AgS_6$ superconductor. Here, we report the effect of In substitution at the Ag site of $La_2O_2Bi_3AgS_6$ compound. $T_c$ decreases with increasing In concentration. The hump at $T^* = 180$ K in the normal state electrical resistivity is not changing clearly with In doping, which is clearly different to the case of Sn doping at the Ag site.

## 2. Experimental Section

The In-doped polycrystalline samples of $La_2O_2Bi_3Ag_{1-x}In_xS_6$ were synthesized by a solid-state reaction method. Powders of $Bi_2O_3$ (99.9%), $La_2S_3$ (99.9%), In (99.99%), and AgO (99.9%) and grains of Bi (99.999%) and S (99.99%) with a nominal composition of $La_2O_2Bi_3Ag_{1-x}In_xS_6$ were ground thoroughly using a pestle and a mortar for mixing, pelletized, sealed in an evacuated quartz tube, and heated at 725 °C for 15 h. For the homogeneity, we reground the obtained samples and heated at 725 °C for 15 h. Finally, we obtained phase-pure samples, and the phase purity was checked by laboratory X-ray diffraction (XRD) with Cu-$K_\alpha$ radiation. The crystal structure parameters were refined using the Rietveld method with RIETAN-FP [24]. A schematic image of the crystal structure was drawn using VESTA [see Figure. 1(c)] [25]. The energy-dispersive X-ray spectroscopy (EDX) on scanning electron microscope TM3030 (Hitachi) was used for analysing actual compositions. The Seebeck coefficient was measured by a four-probe technique on ZEM-3 (Advance RIKO). The electrical resistivity down to $T = 0.4$ K was measured by four probe method using $^3$He probe platform on the Physical Property Measurement System (PPMS: Quantum Design).

## 3. Results and discussion

The room temperature XRD patterns of the In-doped $La_2O_2Bi_3Ag_{1-x}In_xS_6$ ($x$ = 0-0.4) compounds are shown in figure 1(a). These compounds are crystallized in the tetragonal structure with the space group of $P4/nmm$. We present the XRD data up to $x$ = 0.4, which is determined as the solubility limit of In for the Ag site in the $La_2O_2Bi_3Ag_{1-x}In_xS_6$ system. We checked fitting (reliability) parameter $R_{wp}$ through the Rietveld refinement of XRD data of $La_2O_2Bi_3Ag_{1-x}In_xS_6$ compounds. We found that the $R_{wp}$ is lower with keeping the M2 site substituted by In. While, the $R_{wp}$ factor was comparably high when the refinement performed by considering both sites (M1 and M2) substitute by In. The Rietveld analysis suggests that the In ions were substituted at the M2 site. The (103) peak position shifts towards the low angle side with increasing In doping level as shown in figure 1(b). Figure 1(c) shows the schematic unit cell of $La_2O_2Bi_3Ag_{0.9}In_{0.1}S_6$, in which chalcogenide layers of (Bi,Ag,In)S and LaO(Bi,Ag)$S_2$ are alternately stacking. The structural parameters have been determined by the Rietveld fitting of powder XRD pattern. The obtained lattice parameters are shown in figures 1(d, e). The estimated lattice parameters are $a$ = 4.0561(1) Å and $c$ = 19.345(1) Å for $x$ = 0, which become $a$ = 4.052(1) Å and $c$ = 19.32(1) Å for $x$ = 0.1. The lattice parameter $c$ decreases with small amount of In doping for $x$ = 0-0.1, and then continuously increases with increasing $x$ for higher $x$. The lattice parameter a and c are increasing for x>0.1 for the In doped compounds, while the lattice parameter a is decreasing for x> 0.1 for the Sn doped compounds. The lattice parameter c firstly decreases for x<0.2 then slightly increases for x>0.2, the value of c=19.458 (1) Å for the 50% Sn doped compound, we observed the clear positive in plane chemical pressure for the Sn doped $La_2O_2Bi_3Ag_{1-x}Sn_xS_6$ compounds [18]. Comparably, the lattice parameter c become larger for the 40% In doped sample, The lattice parameter $c$ = 19. 647(1) Å for the 40% In doped sample. As well, the continuous increase in lattice parameter $a$ suggests in-plane expansion of the unit cell, which should result the negative chemical pressure effect. The ratio of Bi, Ag and In in the $La_2O_2Bi_3Ag_{1-x}In_xS_6$ ($x$ = 0-0.4) samples were studied by using EDX. To estimate the Bi amount, we tentatively analysed the composition according to the formula of $La_2O_2Bi_{3+y}(Ag_{1-x}In_x)_{1-y}S_6$. The analysed values of $x_{EDX}$ and $y_{EDX}$ are plotted in figures 1(f, g). We found that Bi is slightly deficient in the doped samples with $x$ = 0.1, 0.2, 0.3 and 0.4, but the estimated $y_{EDX}$ was close to zero.

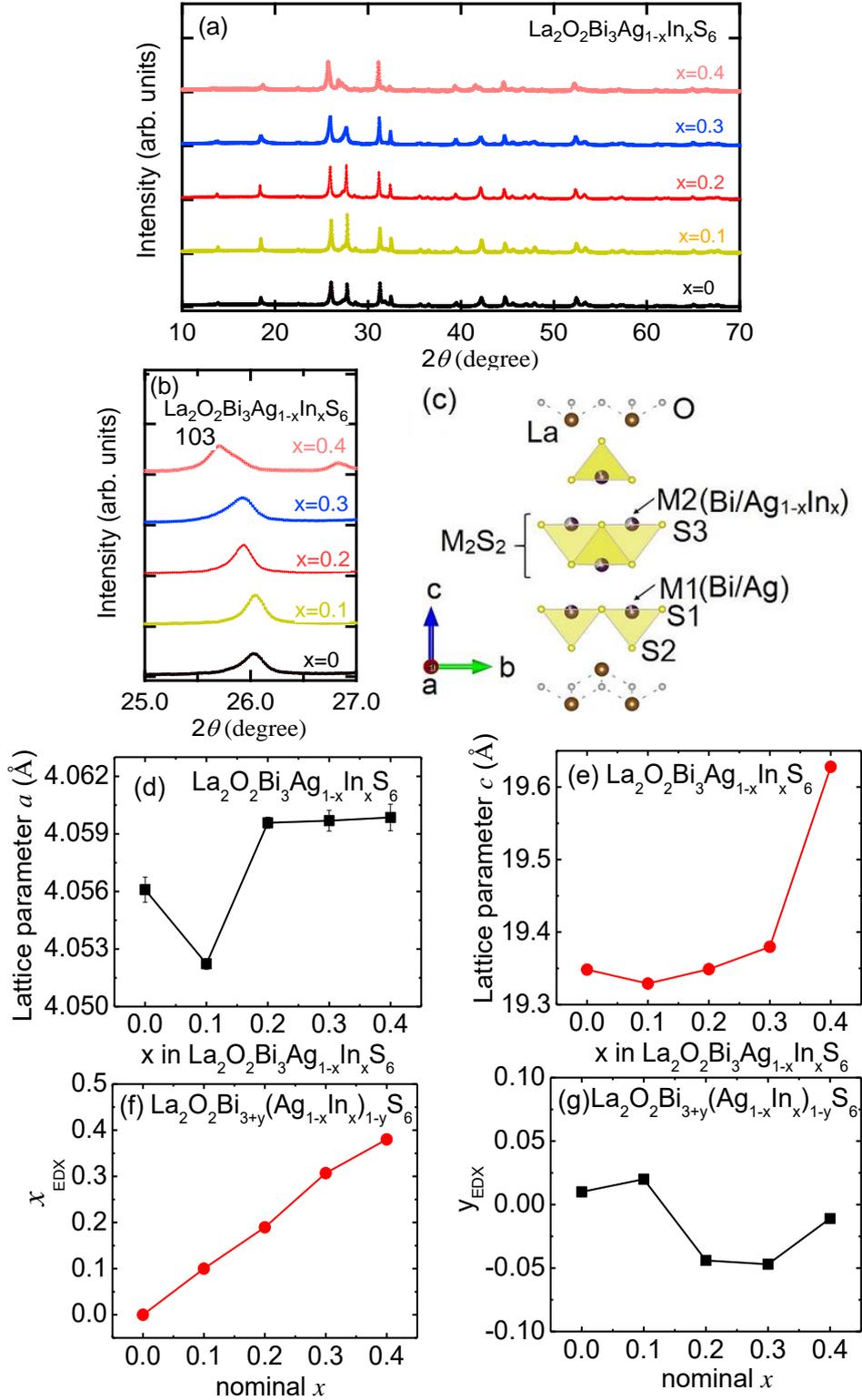

**Figure. 1:** (a) XRD patterns of $La_2O_2Bi_3Ag_{1-x}In_xS_6$ ($x$ = 0-0.4) compounds scanned at room temperature. (b) Zoomed view of XRD patterns near the (103) (Miller indices) peak of tetragonal phase of the $La_2O_2Bi_3AgS_6$ structure. (c) The schematic unit cell of $La_2O_2Bi_3Ag_{0.9}In_{0.1}S_6$. (d, e) Lattice parameters $a$ and $c$ estimated by the Rietveld refinement. (f, g) The nominal composition dependence of $x_{EDX}$ and $y_{EDX}$ in $La_2O_2Bi_{3+y}(Ag_{1-x}In_x)_{1-y}S_6$ obtained by EDX.

Therefore, we assumed the composition of present samples as $La_2O_2Bi_3Ag_{1-x}In_xS_6$; hence, we name the samples using nominal composition of $La_2O_2Bi_3Ag_{1-x}In_xS_6$ in this paper. On the other hand, the In concentration is linearly increasing, which suggests that In was substituted at the Ag site as expected. Because $x_{EDX}$ is almost the same with the nominal x and $y_{EDX}<0.1$, we use $La_2O_2Bi_3Ag_{1-x}In_xS_6$.

The temperature dependences of electrical resistivity $[\rho(T)]$ from 300 to 0.1 K for the In-doped $La_2O_2Bi_3Ag_{1-x}In_xS_6$ ($x = 0, 0.1, 0.2, 0.3$, and $0.4$) are shown in figure 2(a). We observed metallic behaviour with a hump at 180 K in the normal state electrical resistivity for $x = 0, 0.1, 0.2$, and $0.3$. For $x = 0.4$, the normal state $\rho(T)$ shows a semiconducting-type behaviour. The anomaly temperature $T^*$ clearly appeared for $x \leq 0.3$ and does not change with increasing $x$, while $T^*$ for the x=0.4 has been confirmed from second order differentiation of normal state resistivity data. The $\rho(T)$ curves near the superconducting transition are shown in the figure 2(b). The $T_c$ slightly shifts towards the lower temperature side with increasing In concentration in $La_2O_2Bi_3Ag_{1-x}In_xS_6$. The lowest $T_c = 0.4$ K was observed for $x = 0.4$.

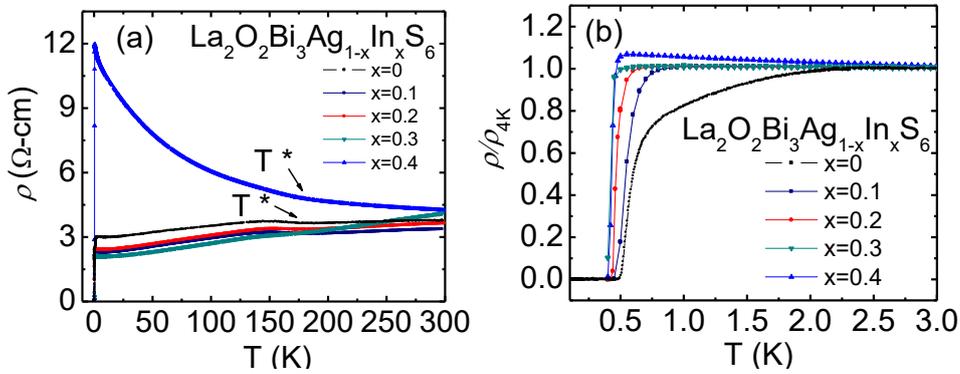

**Figure. 2:** (a) the $\rho(T)$ from 300 K to down to 0.1 K for the $La_2O_2Bi_3Ag_{1-x}In_xS_6$ (x=0-0.4) compounds. A hump structure appearing in the $\rho(T)$ curve is indicated by the arrow at $T = T^*$. (b) The normalized $\rho(T)/\rho(4 K)$ curve in the temperature range 3.0-0.1K.

Figure 3 shows the room temperature Seebeck coefficient for $La_2O_2Bi_3Ag_{1-x}In_xS_6$ ($x = 0, 0.1, 0.2, 0.3$, and $0.4$) samples. For all the samples, Seebeck coefficient is negative, which suggests that the dominant carrier is electron. In addition, we observed no clear change in the Seebeck coefficient by the In doping. The Seebeck coefficient measurements suggest that the In doping is not effective for carrier doping.

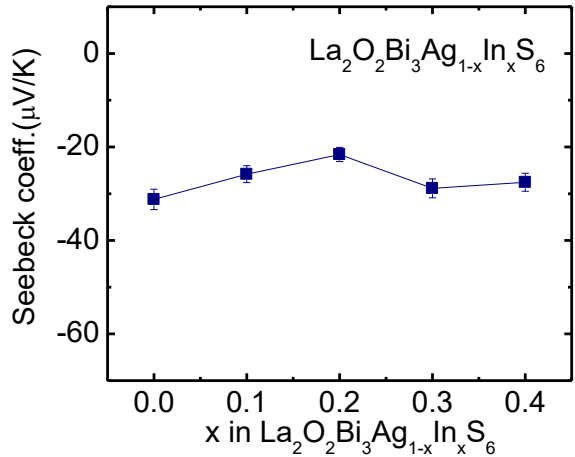

**Figure. 3:** The In concentration (x) dependence of Seebeck coefficient for $La_2O_2Bi_3Ag_{1-x}In_xS_6$ ($x = 0$-$0.4$) at room temperature.

Figure 4 exhibits the $T$ vs $x$ plot (phase diagram) of $La_2O_2Bi_3Ag_{1-x}In_xS_6$, which shows the change in superconducting transition $T_c^{zero}$ with the In doping. The hump in the normal state resistivity ($T^*$) is not changing with increasing In doping. $T_c$ slightly decreases with increasing $x$ in $La_2O_2Bi_3Ag_{1-x}In_xS_6$. From the Seebeck coefficient, In substitution cannot increase carrier density. Therefore, the decrease in $T_c$ for the In-doped samples might be caused by the negative in-plane chemical pressure effect caused by the in-plane expansion of the lattice by In doping. Furthermore, there is the possibility of slight substitution of In at the superconducting BiS plane (as indicated by $y_{EDX} < 0.1$). As reported in Ref. 26, $LaOInS_2$ was synthesized, but the $InS_2$ layer was highly disordered; split model for the In site was needed to refine the XRD data [26]. If small amount of In is substituted for the superconducting BiS plane, introduction of disorder is expected.

Although there is no evidence of charge-density-wave (CDW) ordering at $T^*$ in the $La_2O_2Bi_3AgS_6$ system, one can assume that the CDW transition occurred at $T^*$ from the analogy to the case of $EuFBiS_2$ [27].

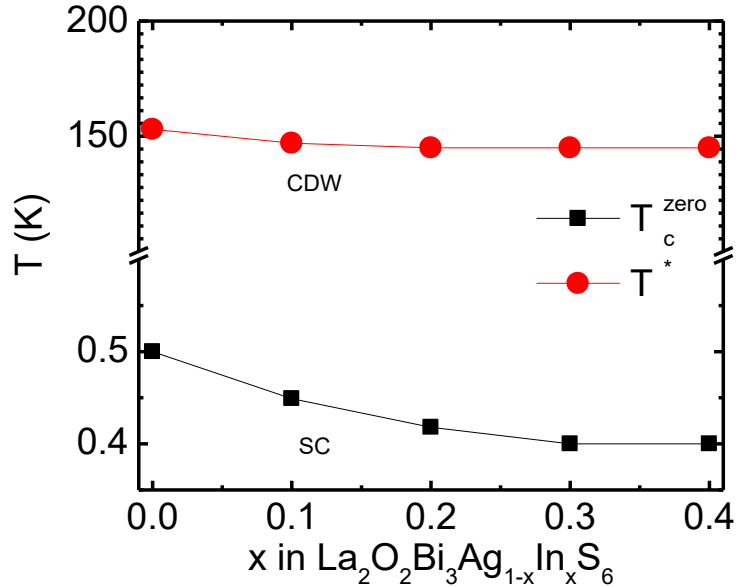

**Figure. 4:** The x dependence of $T_c$ for $La_2O_2Bi_3Ag_{1-x}In_xS_6$ (x=0- 0.5) compounds and $T^*$ scaled on the $x$-$T$ phase diagram for $La_2O_2Bi_3Ag_{1-x}In_xS_6$. CDW and SC symbolises as charge density wave and superconductivity respectively.

Our latest results follow the contemporary scenario, we have observed decrease of the $T^*$ with increasing Sn doping level and enhancement of $T_c$ for the Sn-doped $La_2O_2Bi_3Ag_{1-x}Sn_xS_6$ system [18]. In contrast, $T^*$ is not clearly decreasing or increasing, while $T_c$ is decreasing marginally with increasing In doping. The interplay of CDW and superconducting orders has been reported through the high pressure study on CDW and superconductivity for the $2H$-$NbSe_2$ system [28]. They observed that with increasing pressure an increase in $T_c$ and a decrease in $T_{CDW}$, which shows that CDW phase and superconducting phase are the two competing phases [28]. Therefore, the contrasting evolution of superconductivity in the Sn-doped and In-doped systems should be related to the robustness of CDW ordering to the substitutions. On the basis of the discussion above, we propose that the suppression of superconductivity in In-doped $La_2O_2Bi_3Ag_{1-x}In_xS_6$ is caused by negative in-plane chemical pressure effect, introduction of disorder at the superconducting BiS plane by substitution of small amount of In. The EDX data show small amount of Bi deficiency, which suggests that the Bi site is slightly substituted by In. The decrease in $T_c$ might be associated with the Bi site substitution by small amount of In. The Seebeck coefficient data suggest that the carrier concentration does not change sufficiently via In doping, which indicates that the CDW transition is robust against the In substitution. To precisely determine the doping site and local structure parameters, synchrotron X-ray diffraction experiment is needed.

## 4. Conclusion

We have investigated the substitution effect of Indium (In) at Ag site in $La_2O_2Bi_3AgS_6$. The $T_c$ decreased with increasing In doping, and $T_c^{zero}$ was 0.4 K for $x = 0.4$. The hump in the normal state resistivity at anomaly temperature ($T^* \sim 180$ K) remained unchanged with In doping up to $x = 0.3$. The $x = 0.4$ sample showed semiconducting-like behaviour with having CDW anomaly. The suppression of $T_c$ in the In-doped $La_2O_2Bi_3Ag_{1-x}In_xS_6$ system is caused by negative in-plane chemical pressure effect, introduction of disorder at the superconducting BiS plane by substitution of small amount of In. The robustness of the CDW ordering to the In substitution also suggests that the carrier concentration does not change sufficiently.

**Acknowledgments**

This work was partly supported by Grants-in-Aid for Scientific Research (KAKENHI) (Grant Nos. 15H05886, 15H05884, 16H04493, 17K19058, 16K05454, and 15H03693).